\documentclass[conference]{IEEEtran}
\IEEEoverridecommandlockouts

\usepackage{cite}
\usepackage{amsmath,amssymb,amsfonts}
\usepackage{algorithmic}
\usepackage{graphicx}
\usepackage{textcomp}
\usepackage{xcolor}
\def\BibTeX{{\rm B\kern-.05em{\sc i\kern-.025em b}\kern-.08em
    T\kern-.1667em\lower.7ex\hbox{E}\kern-.125emX}}

\usepackage{amssymb}
\usepackage{pifont}
\usepackage{multirow}
\usepackage{tipa}
\newcommand{\cmark}{\ding{51}}%
\newcommand{\xmark}{\ding{55}}%

\usepackage[T1]{fontenc}

\def\ninept{\def\baselinestretch{.95}\let\normalsize\small\normalsize}

\begin{document}
\small

\title{Improving Cross-Lingual Phonetic Representation of Low-Resource Languages Through Language Similarity Analysis
 \thanks{This work was supported by the National Research Foundation of Korea grant funded by the Korea government (MSIT). (No. 2021R1A2C1014044)}
}

\author{\IEEEauthorblockN{Minu Kim}
\IEEEauthorblockA{\textit{School of Electrical Engineering} \\
\textit{KAIST}\\
Daejeon, Republic of Korea \\
minus@kaist.ac.kr}
\and
\IEEEauthorblockN{Kangwook Jang}
\IEEEauthorblockA{\textit{School of Electrical Engineering} \\
\textit{KAIST}\\
Daejeon, Republic of Korea \\
dnrrkdwkd12@kaist.ac.kr}
\and
\IEEEauthorblockN{Hoirin Kim}
\IEEEauthorblockA{\textit{School of Electrical Engineering} \\
\textit{KAIST}\\
Daejeon, Republic of Korea \\
hoirkim@kaist.ac.kr}
}

\maketitle
%
%
\begin{abstract}
This paper examines how linguistic similarity affects cross-lingual phonetic representation in speech processing for low-resource languages, emphasizing effective source language selection. Previous cross-lingual research has used various source languages to enhance performance for the target low-resource language without thorough consideration of selection. Our study stands out by providing an in-depth analysis of language selection, supported by a practical approach to assess phonetic proximity among multiple language families. We investigate how within-family similarity impacts performance in multilingual training, which aids in understanding language dynamics. We also evaluate the effect of using phonologically similar languages, regardless of family. For the phoneme recognition task, utilizing phonologically similar languages consistently achieves a relative improvement of 55.6\% over monolingual training, even surpassing the performance of a large-scale self-supervised learning model. Multilingual training within the same language family demonstrates that higher phonological similarity enhances performance, while lower similarity results in degraded performance compared to monolingual training.
\end{abstract}


\section{Introduction}
\label{sec:intro}

There are approximately 7,000 languages in the world, but the number studied in the field of speech processing is limited.
The primary reason for this is the scarcity of data, leading to the classification of these languages as \textit{low resource languages} \cite{shamsfard2019challenges}.
Given the limited data available for low resource languages, training models exclusively on these languages poses significant challenges.
Some research \cite{vu2012investigation, lam2023multilingual} has shown that multilingual processing can enhance recognition for new languages by leveraging knowledge from multiple sources. However, effective source language selection remains underexplored. This study follows a similar approach by incorporating multiple languages alongside the target language for training. Furthermore, this work proposes a strategy for optimal source language selection in cross-lingual training for low-resource speech recognition.

On the other hand, research in Natural Language Processing (NLP), particularly in text-based tasks, has focused on selecting languages for cross-lingual tasks based on linguistic similarity. Research like \cite{philippy2023towards} evaluates how similarity impacts model performance, while the study by \cite{lin2019choosing} ranks languages for training based on linguistic criteria. In most of cross-lingual speech recognition studies, however, source languages are often chosen randomly \cite{dalmia2018sequence, hou2021exploiting, wang2020improving} or taken from the same language family without much consideration \cite{lu2013cross, yu2019cross}. Though not their main focus, some studies have attempted to demonstrate the impact of source languages. For example, the study by \cite{zellou2024linguistic} examines International Phonetic Alphabet (IPA) transcription based on similarity, and the research conducted by \cite{plahl2011cross} shows that closely related languages perform better in cross-linguistic contexts. Other works \cite{tachbelie2020development, krishna2021using, hu2024cam} support using phonologically similar languages. Meanwhile, conflicting views on family-based learning exist: some report training within the same language family improves performance \cite{huang2013cross}, while others find training with other language family data can be effective \cite{mussakhojayeva2023multilingual}.

However, existing research on cross-lingual speech recognition for low-resource languages has not mainly focused on the selection of languages itself, resulting in a lack of detailed analysis of its effects and the absence of specific metrics. This study aims to explore the reasoning behind these seemingly contradictory findings in cross-lingual speech recognition and seeks to determine how to effectively implement cross-lingual approaches to enhance performance in low-resource settings. By applying cross-lingual studies across diverse language families, this research will provide insights into the dynamics of language similarity and its impact on speech recognition.

As noted by \cite{bentz2016comparison}, languages can be evaluated through corpus-based and typology-based approaches. In the corpus-based similarity assessment, we convert the text in the speech processing data into IPA and analyze the frequency distribution of phonemes \cite{mines1978frequency}. The languages are evaluated as more similar if their phoneme distributions are alike. In the typology-based similarity assessment, we examine how similar the typological features are using the Grambank dataset \cite{skirgaard2023grambank}, which numerically records the typological characteristics of languages. In this study, we primarily utilize corpus-based similarity assessment, while typology-based similarity evaluation is employed as a supplementary method to examine how well it aligns with the similarity evaluation of data within the same language family.

We compare three language families—Indo-Iranian (a branch of Indo-European), Turkic, and Afro-Asiatic—using both methods to assess the degree of similarity among languages within each family. As languages in the same family phylogenetically share the same root (ancestor), it can be a starting point for cross-lingual approach. Based on this analysis, we will examine the impact of language similarity in family-based multilingual learning. Additionally, we will explore how training language data selected based on phonological distribution similarity regardless of language family (obtained from corpus-based methods) affects cross-lingual learning. As we train a universal symbol system based on the IPA, no language-specific fine-tuning is conducted during multilingual training. Instead, the model is trained to produce the corresponding IPA when inputting speech data from various languages. For both monolingual and multilingual training, we utilize a Conformer-based model \cite{gulati2020conformer}.

The results of our analysis reveal a clear trend: the performance of our model improves significantly when there is greater phonological similarity between the source and target languages within the same language family. This performance boost is particularly evident when the phonological characteristics of these languages are closely aligned. Conversely, with lower phonological similarity, performance drops, and in some cases, it can be worse than monolingual training, where the language pairs lack phonological advantages. Leveraging languages with high phonological similarity (regardless of family) yields a relative improvement of 55.6\% over the monolingual approach, notably surpassing the performance of a large-scale self-supervised learning (SSL) model\,\cite{conneau2020unsupervised}. This advantage is particularly notable given that fine-tuning the SSL model requires extensive pre-training data and significantly larger parameters (300M vs. 43M), which underscores the effectiveness of our source selection approach.

\section{Analysis Methods}

\begin{figure}[!t]
    \centering
    \includegraphics[width=\columnwidth]{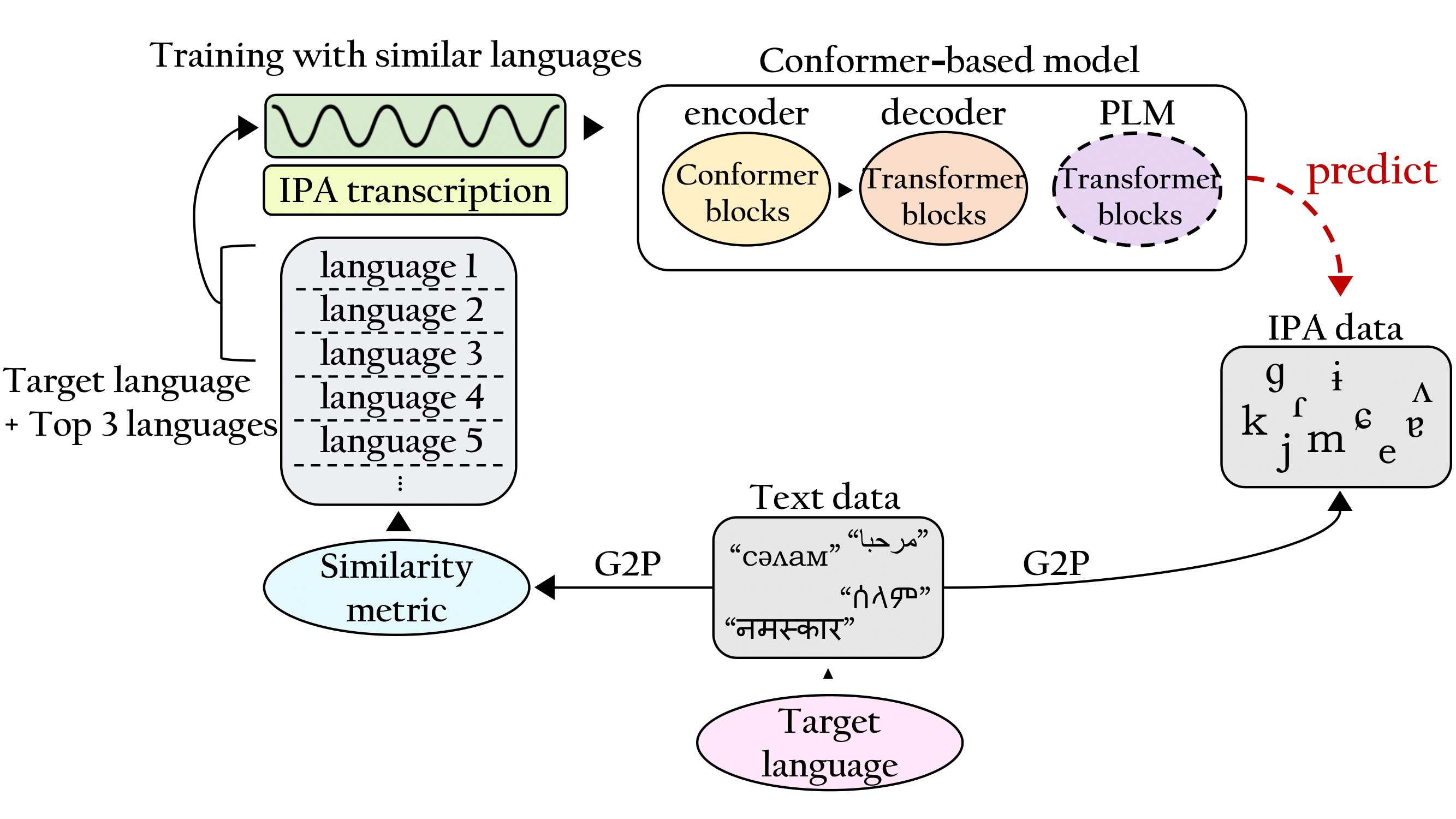}
    \vspace{-18pt}
    \caption{The Conformer-based model is trained to infer IPA sequences for each low-resource language, with the best results achieved using the top 3 most phonologically similar languages (\textit{i.e.}\,$|\mathcal{S}|=3$ in Equation. (3a)) with the target language. The Phoneme-level Language Model (PLM) may or may not be used in decoding depending on experimental conditions.}
    \label{fig:overall}
\end{figure}

This chapter contains methods for constructing an appropriate training set to learn phonetic representations from the speech of the target language. The overall framework is illustrated in Fig. \ref{fig:overall}.

\subsection{Grapheme-to-Phoneme Model}
\label{sec:g2p-model}

For low-resource languages, especially those at risk of extinction, IPA transcription is crucial as it captures the phonological essence of the language \cite{michailovsky2014documenting}, providing a key resource for research, revitalization, and education. Traditionally, human listeners manually transcribe spoken language into IPA, which is labor-intensive and time-consuming \cite{wittenburg2002methods}.
There have been efforts to automate this process \cite{xu2021simple, yan2021differentiable, taguchi2023universal, bharadwaj2016phonologically}, and our study builds on those attempts.
We transcribe text data into IPA using a Grapheme-to-Phoneme (G2P) model.
\subsection{Similarity Metric}
\label{sec:sim-measure}

In this section, we suggest two language similarity metrics: corpus-based and typology-based.
Each methodology is used to evaluate the similarity among languages within a family, and specifically, the corpus-based approach is employed to select the phonetically most similar languages regardless of their language family.

\textbf{\textit{Corpus-based approach.}} \quad We analyze unigram phoneme sequences by examining the frequencies of individual phonemes and measuring similarity using cosine similarity between phoneme distribution vectors \( \mathbf{p}_A \) and \( \mathbf{p}_B \) for languages \(A\) and \(B\). These vectors are derived from the text information of speech processing datasets, which is converted into phoneme sequences through a G2P model. The phonological similarity, calculated as shown in Equation (1), helps guide dataset selection for cross-lingual learning.

\begin{equation}
    \text{cos}(\mathbf{p}_A, \mathbf{p}_B) = \frac{\mathbf{p}_A \cdot \mathbf{p}_B}{\parallel\mathbf{p}_A\parallel \parallel\mathbf{p}_B\parallel}
    \label{equ:lang_sim}
\end{equation}

A similarity matrix \( S \in \mathbb{R}^{N \times N} \) on $N$ languages is constructed, with each cell representing the cosine similarity between phoneme distributions of two languages. Principal Component Analysis (PCA) is then applied to \( S \), resulting in a 2D projection matrix \( S_{\text{PCA}} \in \mathbb{R}^{N \times 2} \) that helps visualize complex relationships while preserving the most significant variance among the languages. Each language is represented by coordinates \( (x_i, y_i) \), with proximity indicating phonetic similarity. Kernel Density Estimation (KDE) \cite{wkeglarczyk2018kernel} is then used to create smooth contours around language clusters in the PCA-reduced space, visualizing language distribution and phonological similarity. The density \( \hat{f}(x, y) \) at point \( (x, y) \) is estimated using the Gaussian kernel function \( K(\cdot) \), based on coordinates \( (x_i, y_i) \) of the \( i \)-th language as shown in Equation (2a). Densities for each language family are calculated separately, with $N_k$ indicating the number of languages in family $k$, and weights $w_i$ accounting for relative recording time. Bandwidth parameters $h_x$ and $h_y$ are set using Silverman's rule of thumb \cite{andersson2014improving}. Each language family's contour \( C_k \) is computed at a density level of $c=0.1$ as in Equation (2b).

\vspace{-10pt}
\begin{align}
\hat{f}(x, y) &= \frac{1}{N_k h_x h_y} \sum_{i=1}^{N_k} w_i K\left( \frac{x - x_i}{h_x} \right) K\left( \frac{y - y_i}{h_y} \right) \tag{2a}\\
C_k &= \left\{ (x, y)  \ \bigg| \ \hat{f}(x, y) = c \right\}, \ k = \text{language family} \tag{2b}
\end{align}

\textbf{\textit{Typology-based approach.}} \quad Prior works\,\cite{bentz2016comparison, skirgaard2023grambank} have shown that when typological descriptions are represented in binary form and PCA is applied, the first two principal components capture grammatical complexity. Similar to the previous research above, we perform PCA on Grambank data to measure this complexity. Grambank \cite{skirgaard2023grambank} is a comprehensive database of grammatical features across languages, including word order, verb conjugation patterns, and specific grammatical markers. Although Grambank is dense compared to the World Atlas of Language Structures (WALS) \cite{haspelmath2008world} dataset, it still has missing values. Since PCA requires complete data, the preceding study \cite{skirgaard2023grambank} uses random forest to predict these missing values before PCA. We incorporate their predictions in our analysis and use it to show how languages from the same family are distributed in the principal component space.
\subsection{Phonetic Transcription Model}
\label{sec:sr-model}

In this study, an encoder-decoder architecture generates phonetic representations through speech processing, trained on either a single or multiple languages based on the environment. Equation (3a) shows the selection process for the top 3 languages most phonetically similar to the target low-resource language. These languages are chosen from the full language set \( \mathcal{L} \) and used to form the training set \( \mathcal{S} \). The text data are converted to IPA using a G2P model and utilized for training as shown in Equation (3b). Here, $\phi$ represents a phoneme sequence obtained by converting the word W using a G2P model, where $\phi_j$ is an individual phoneme in the sequence $\phi$. Additionally, $\mathcal{P}_{\text{language}_i}$ is a phoneme set of language $i$, and $X$ denotes the speech data.

\vspace{-10pt}
\begin{align}
\mathcal{S} &= \mathop{\text{argmax}}_{\mathcal{L}_j \in \mathcal{L}, j \neq \text{target}} \Big( \cos(\mathbf{p}_{\text{target}}, \mathbf{p}_j) \Big), \ |\mathcal{S}| = 3 \tag{3a}\\
\phi &= \mathcal{G_{\text{2}}P}(\text{W}), \ \text{where} \ \forall \phi_j \in \phi, \ \phi_j \in \bigcup_{\mathcal{L}_i \in \mathcal{S}} \mathcal{P}_{\text{language}_i} \tag{3b}\\
\phi^{\ast}&=\mathop{\text{argmax}}_{\phi} P(\phi | X) =\mathop{\text{argmax}}_{\phi} P(X \vert \phi)P(\phi) \tag{3c}
\end{align}

\section{Experimental Setups}

\textit{\textbf{Dataset.}} \quad
Common Voice\,\cite{ardila2019common} version 18.0 dataset is used for training phonetic transcription model.
Word representations are converted to IPA using a G2P model, such as Epitran\,\cite{mortensen2018epitran}, CommonVoiceUtils\footnote{https://github.com/ftyers/commonvoice-utils}, and Gruut\footnote{https://github.com/rhasspy/gruut}.
We perform post-adjustments on the data, such as omitting voice quality symbols\,(VoQS), removing certain diacritics like stress indicators, and merging phoneme pairs\,(e.g., \textipa{[/S/, /s\super j/] or [/Z/, /z\super j/]}) that are not significantly distinguished in phonology.
Only validated train utterances, which are approved by at least two of three evaluators, are used for training.
Table \ref{tab:language_data} describes the languages and corresponding families used in our experiments.
Similar to the use of 10 hours of low-resource data in the previous study \cite{san2024predicting}, we define languages with less than 15 hours of recorded data as low-resource languages and utilize them for inference.

\begin{table}[!t]
\caption{We utilize three language families, comprising a total of 22 languages, for our IPA transcription task. We define low-resource languages as those with less than 15 hours of validated recording data.}
\label{tab:language_data}
\centering
\begin{tabular}{cccc}
     \hline
     \multirow{2}{*}{\textbf{Family}} & \multirow{2}{*}{\textbf{Language}} & \multicolumn{1}{c}{\textbf{Recording time}} & \multirow{2}{*}{\textbf{Low-resource}} \\ 
     & & \multicolumn{1}{c}{\textbf{(in hours)}} & \\ \hline\hline
     \multirow{6}{*}{\centering\textbf{Indo-Iranian}} 
     & Punjabi & 2.29 & \cmark \\ \cline{2-4}
     & Hindi & 14.71 & \cmark \\ \cline{2-4}
     & Bengali & 53.62 & \xmark \\ \cline{2-4}
     & Urdu & 63.69 & \xmark \\ \cline{2-4}
     & Kurdish & 67.83 & \xmark \\ \cline{2-4}
     & Persian & 365.64 & \xmark\\ \hline
     \multirow{9}{*}{\centering\textbf{Turkic}} 
     & Azeri & 0.33 & \cmark \\ \cline{2-4}
     & Kazakh & 2.15 & \cmark \\ \cline{2-4}
     & Turkmen & 2.75 & \cmark \\ \cline{2-4}
     & Sakha & 8.35 & \cmark\\ \cline{2-4}
     & Tatar & 30.66 & \xmark \\ \cline{2-4}
     & Uzbek & 99.81 & \xmark \\ \cline{2-4}
     & Turkish & 120.05 & \xmark \\ \cline{2-4}
     & Uyghur & 232.67 & \xmark \\ \cline{2-4}
     & Bashkir & 258.04 & \xmark \\ \cline{2-4} \hline
     \multirow{7}{*}{\centering\textbf{Afro-Asiatic}} 
     & Tigrinya & 0.03 & \cmark \\ \cline{2-4}
     & Tigre & 1.12 & \cmark \\ \cline{2-4}
     & Amharic & 1.60 & \cmark \\ \cline{2-4}
     & Hausa & 3.95 & \cmark \\ \cline{2-4}
     & Maltese & 8.64 & \cmark \\ \cline{2-4}
     & Arabic & 90.44 & \xmark \\ \cline{2-4}
     & Kabyle & 567.32 & \xmark\\ \hline
\end{tabular}
\end{table}
\textit{\textbf{IPA Transcription Model.}} \quad
We exploit an encoder-decoder structure based on Conformer\,\cite{gulati2020conformer} and Transformer\,\cite{vaswani2017attention}, for the IPA transcription model.
The encoder and decoder consist 12 and 6 layers, respectively. The detailed hyperparameters follow the CVC recipe \cite{mehta2018espnet}.
A Phoneme-level Language Model (PLM)\,\cite{dalmia2019phoneme} is integrated into the decoder side, utilizing 16 Transformer layers to learn global sound representations and align phonemes across languages.

\section{Experiments}

\subsection{Language Similarity}
\label{sec:lang-sim}

\begin{figure}[!t]
    \centerline{\includegraphics[width=0.7\columnwidth]{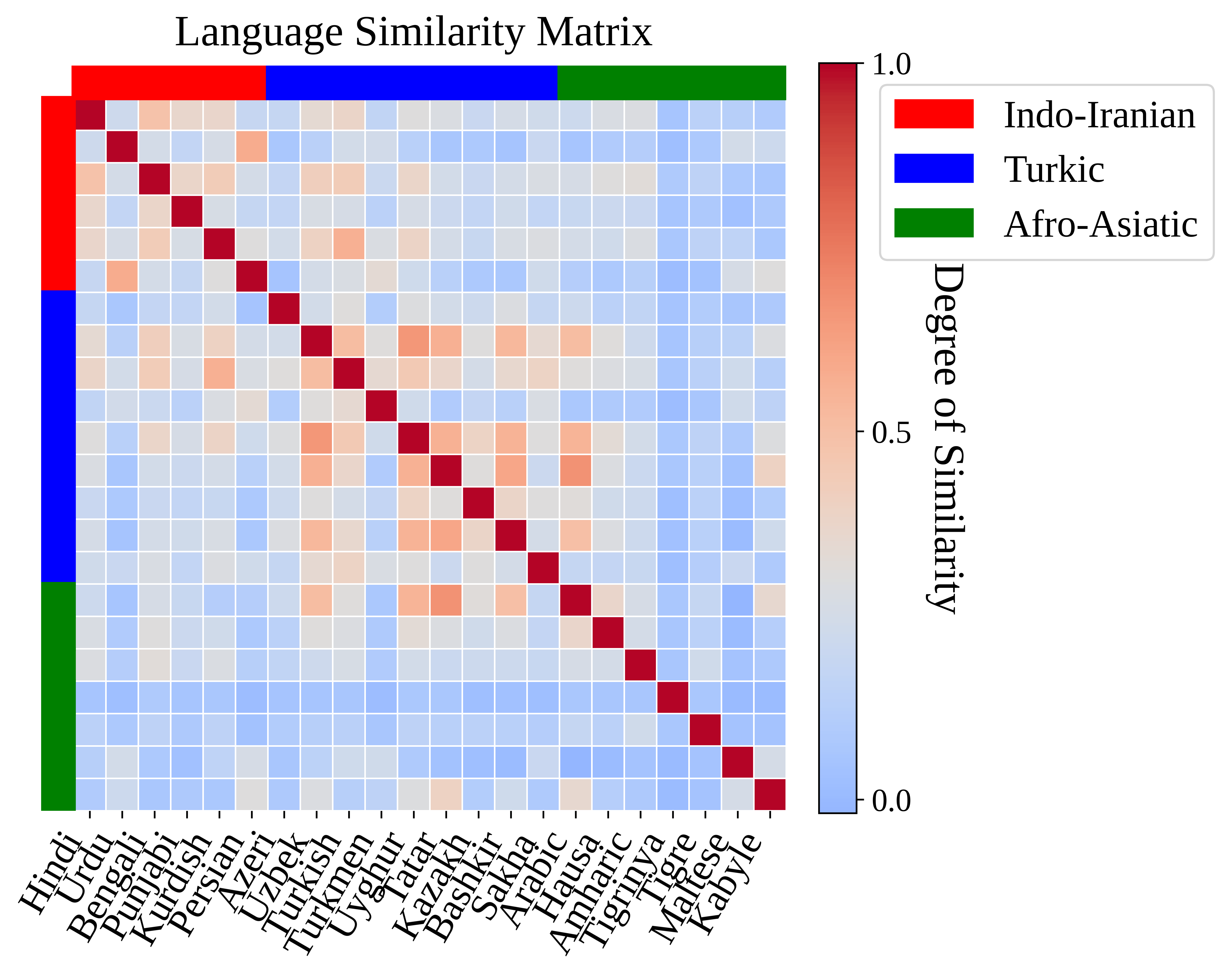}}
    \vspace{-5pt}
    \caption{The language similarity matrix measures cosine similarity between the 22 languages based on phoneme distribution. It provides pairwise similarity for each language, with each row representing its embedding.}
    \label{fig:lang_mat}
    \vspace{-5pt}
\end{figure}

\begin{figure}[!t]
    \centerline{\includegraphics[width=\columnwidth]{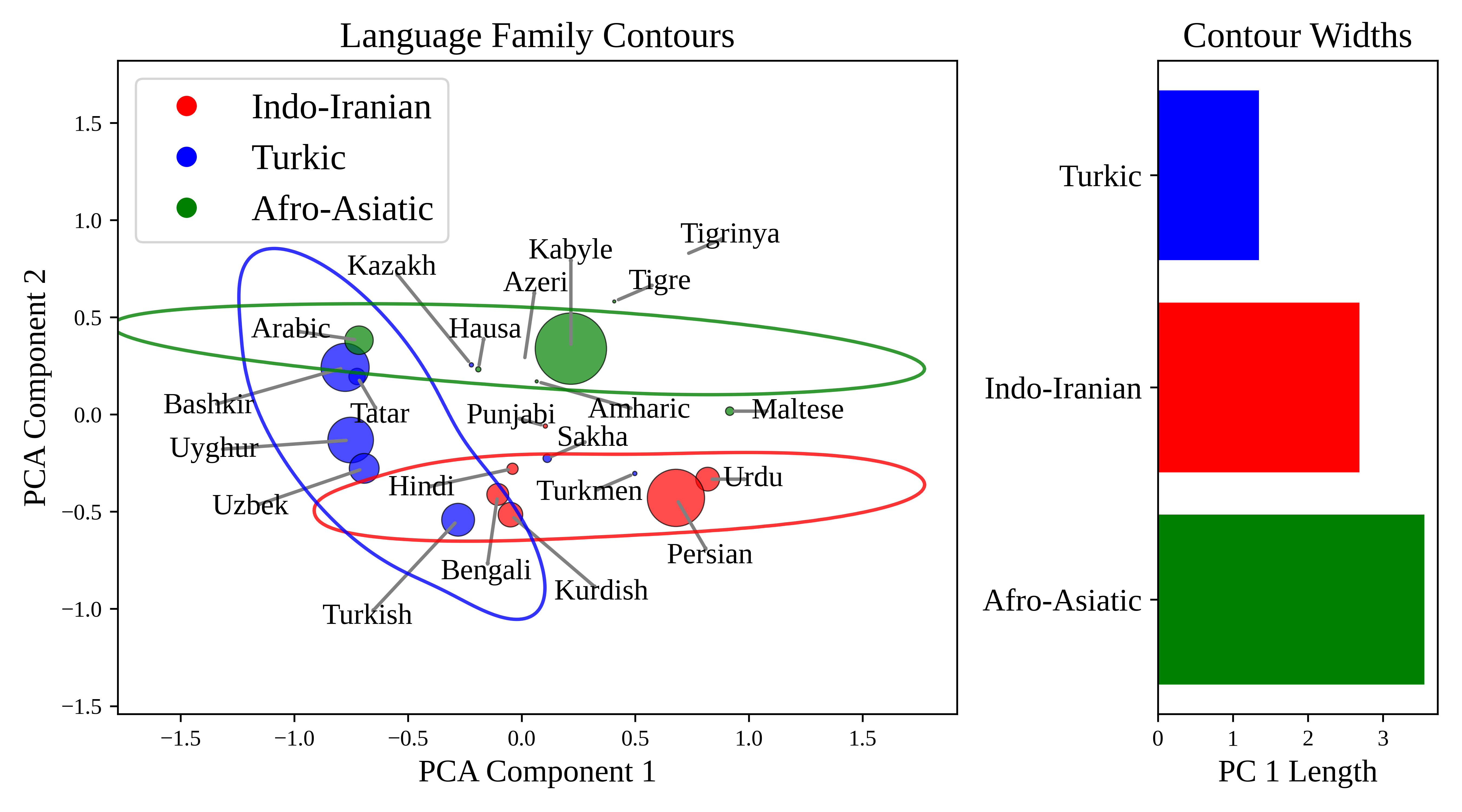}}
    \vspace{-5pt}
    \caption{(Left) Language family contours, based on phoneme distribution similarities, illustrate phonological relationships. (Right) Contour widths represent the first PCA component, showing the highest similarity among Turkic languages compared to other families.}
    \label{fig:corpus_sim} 
\end{figure}

\textbf{\textit{Corpus-based approach.}} \quad 
Fig. \ref{fig:lang_mat} depicts the similarity matrix of language vectors based on unigram phoneme sequences. From this, we can see that there are language pairs that appear to have high similarity within the same language family, but there are also those that do not.
Language family contours on left side of Fig. \ref{fig:corpus_sim} are generated for each language family--Indo-Iranian, Turkic, and Afro-Asiatic--showing the spatial distribution of language points in a 2D PCA space.
The mapping illustrates the concept of \textit{Sprachbund}\,\cite{schaller1997roman}, where geographically close languages become similar even if they are not from the same family.
Arabic, an Afro-Asiatic language, appears within the Turkic contour due to its historical and cultural influence on Turkic languages, leading to shared expressions and loanwords\,\cite{johanson2021history} among languages.
Similarly, Kurdish, an Indo-Iranian language spoken in Turkey where Turkish is predominant\,\cite{kreyenbroek2005kurdish}, is placed within the Turkic contour, resulting in Turkish and Kurdish being mapped closely together.
The PCA results reveal that Turkic languages are the most similar to each other, while Indo-Iranian and Afro-Asiatic families are more distanced, particularly along the most significant principal component\,(refer to right side of Fig. \ref{fig:corpus_sim}).

\begin{table*}[!t]
\centering
\footnotesize
\caption{PER\,(\%) for low-resource languages by model type, source selection type, and whether PLM is used or not.
We train the Conformer model with only target language (monolingual), or adding languages selected by various strategies (family, all, corpus sim.\,(similarity)).
We also compare our results with SSL model (XLSR-53 \cite{conneau2020unsupervised}) fine-tuned on the target low-resource language.
Note that XLSR-53 requires 56k hours of various languages to pretrain.}
\label{tab:main_results}
\addtolength{\tabcolsep}{-2pt}
\renewcommand{\arraystretch}{1.1}
\resizebox{\textwidth}{!}{
\begin{tabular}{ccc|cc|cccc|ccccc|c}
\hline
\multicolumn{3}{c|}{\textbf{Models}}
 & \multicolumn{2}{c|}{\textbf{Indo-Iranian}} & \multicolumn{4}{c|}{\textbf{Turkic}} & \multicolumn{5}{c|}{\textbf{Afro-Asiatic}} & \multicolumn{1}{c}{} \\ 
  \textbf{Backbone} & \textbf{Source selection} & \textbf{PLM} & \textbf{Punjabi} & \textbf{Hindi} & \textbf{Azeri} & \textbf{Kazakh} & \textbf{Turkmen} & \textbf{Sakha} & \textbf{Tigrinya} & \textbf{Tigre} & \textbf{Amharic} & \textbf{Hausa} & \textbf{Maltese} & \textbf{Average} \\ \hline\hline

\textbf{Conformer} & monolingual & & 67.7 & 23.9 & 83.9 & 40.3 & 43.6 & 22.6 & 81.6 & 74.5 & 82.9 & 40.8 & 72.2 & 57.6 \\ 
\textbf{Conformer} & family & & 79.0 & 80.1 & 20.6 & 21.6 & 27.6 & 39.0 & 84.1 & 82.4 & 86.2 & 82.9 & 81.9 & 62.3  \\  
\textbf{Conformer} & all & & 41.2 & 41.8 & 22.6 & 23.0 & 28.6 & 26.8 & 46.4 & 29.6 & 36.6 & 41.9 & 29.8 & 33.5\\
\textbf{XLSR-53} \cite{conneau2020unsupervised} & - & & 25.1 & 22.8 & 36.4 & 17.8 & 25.6 & 19.5 & 98.7 & 23.3 & 31.1 & 35.7 & 20.3 & 32.4 \\ \hline
\textbf{Conformer} & corpus sim.\,\textbf{(Ours)} & & 38.4 & 23.5 & \textbf{19.0} & 19.3 & \textbf{20.2} & 19.8 & 50.2 & \textbf{21.6} & \textbf{29.0} & 36.4 & 21.5 & \textbf{27.2} \\ \hline
\hline 
\textbf{Conformer} & monolingual &\cmark & 67.6 & 23.2 & 82.5 & 40.0 & 43.2 & 22.2 & 75.3 & 75.0 & 83.8 & 40.1 & 74.2 & 57.0\\
\textbf{Conformer} & family & \cmark & 80.2 & 80.4 & 20.8 & 18.4 & 24.2 & 39.7 & 84.5 & 82.6 & 86.2 & 83.4 & 82.5 & 62.1 \\
\textbf{Conformer} & all & \cmark & 42.5 & 43.5 & 24.8 & 21.0 & 26.7 & 25.3 & 52.3 & 29.2 & 35.1 & 43.7 & 25.8 & 33.6\\ 
\textbf{XLSR-53} \cite{conneau2020unsupervised} & - & \cmark & 24.9 & 22.4 & 36.2 & 17.7 & 25.3 & 19.3 & 98.0 & 23.3 & 31.3 & 35.1 & 20.5 & 32.2 \\ \hline
\textbf{Conformer} & corpus sim.\,\textbf{(Ours)} & \cmark & 36.8 & 22.6 & \textbf{18.1} & \textbf{17.4} & \textbf{18.4} & \textbf{19.2} & \textbf{45.3} & \textbf{19.1} & \textbf{26.2} & 35.3 & \textbf{19.8} & \textbf{25.3} \\ \hline \hline
\end{tabular}}
\end{table*}

\begin{figure}[!t]
    \centerline{\includegraphics[width=0.9\columnwidth]{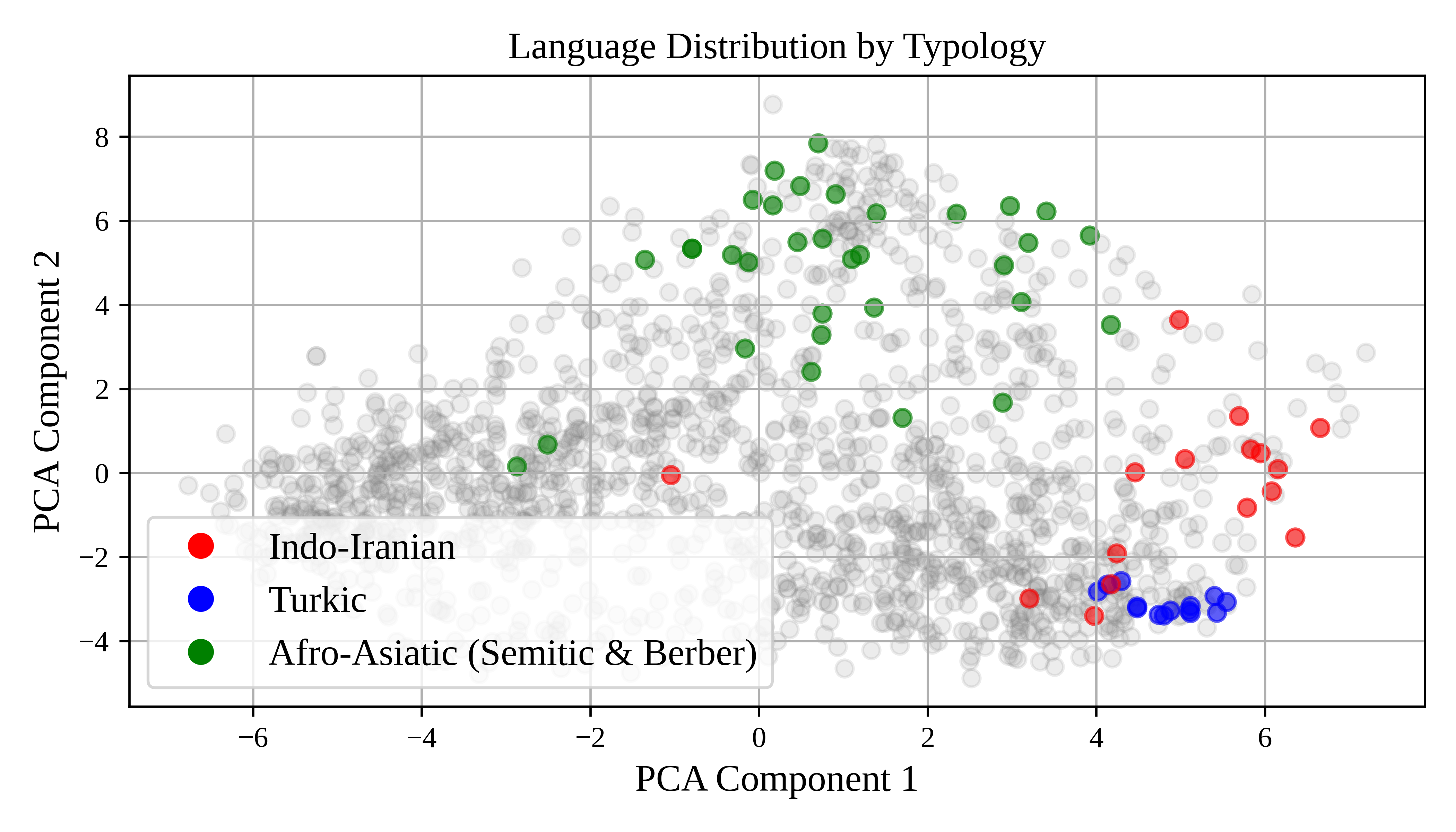}}
    \caption{Typological language distribution reflects similarity between languages and supports the corpus-based similarity assessment. The Turkic family shows the highest similarity, while other families exhibit lower similarity.}
    \label{fig:typ_sim} 
\end{figure}

\textbf{\textit{Typology-based approach.}} \quad The results of PCA on the binary features of all languages in Grambank are shown in Fig. \ref{fig:typ_sim}. Gray points represent the distribution of all languages, while colored points indicate those from the families used for training. For the Afro-Asiatic family, only the Semitic and Berber branches used for training are highlighted. This plot assesses the distribution to reflect the similarity among language families, encompassing all languages in the branches as fully as possible. The typology-based analysis reinforces our earlier findings, revealing that Turkic languages exhibit the highest similarity among the evaluated families, while the others show lower similarity. The results of the corpus-based language similarity assessment are supported by the typology-based evaluation, strengthening our conclusions. Through both corpus-based and typology-based approaches, we have provided comprehensive assessments of language similarity within and across the selected language families.
\subsection{Model Comparisons}
\label{sec:model-compar}

\textbf{\textit{The effects of phylogenetics.}} \quad
Phylogenetic linguistics involves identifying languages in the same family, and phylogenetic-based\,(family-based) multilingual learning can be a starting point for this research.
Phoneme Error Rate\,(PER)\,\cite{he2021performance} of IPA transcription is measured for low-resource languages shown in Table\,\ref{tab:main_results}.
For Turkic family, where languages are similar within the family, family-based training shows significant improvements over monolingual training.
However, adding data from other families\,(denoted as “all” in Table\,\ref{tab:main_results}) generally does not surpass the performance of family-based multilingual training for Turkic languages, which already benefit from their inherent similarities.
Sakha, on the other hand, does not improve with family-based training because it belongs exclusively to the Siberian branch, while the others are from the Oghuz or Chagatai branches.
Furthermore, Sakha is used mainly in Siberia, whereas other Turkic languages are used in Central Asia.
In families where languages are dissimilar, such as Indo-Iranian or Afro-Asiatic, family-based multilingual training may underperform monolingual models. In this case, incorporating data from languages of other families can help bridge the disconnect between the languages.

\begin{figure}[!t]
    \centerline{\includegraphics[width=\columnwidth]{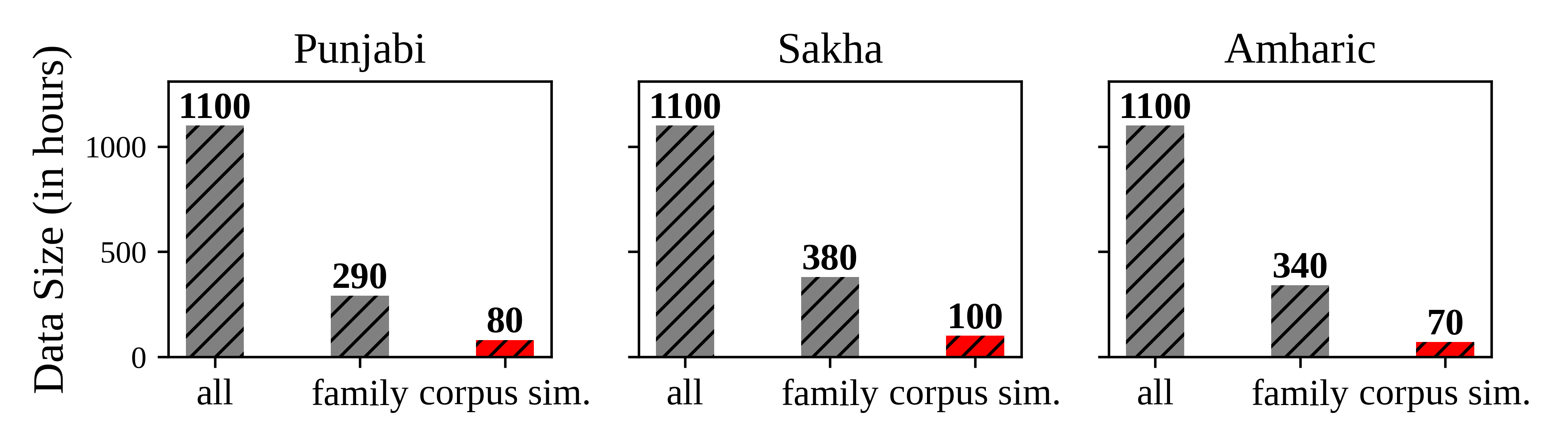}}
    \caption{The training data amounts for three representative languages from each language family are presented. This includes the recorded data size of using all 22 languages, family-based languages, and the three closest languages. The overall trends are similar for other languages as well.}
    \label{fig:data_time} 
\end{figure}

\textbf{\textit{The effects of phonology.}} \quad
Training with the top three phonologically closest languages, as identified in Table\,\ref{tab:language_data}, consistently outperforms monolingual training. Our method not only exceeds the performance of approaches using all data from the language set but also surpasses multilingual training within the same language families. This suggests that phonological similarity improves cross-lingual learning in phonetic representation, emphasizing the importance of configuring the training set to the target language.
As illustrated in Fig. \ref{fig:data_time}, our metric requires the least data time for multilingual approaches, while family-based training demands 3 to 4 times more, and using all 22 languages requires over 10 times. Yet, extensive datasets do not outperform training sets with fewer but more similar languages, highlighting the effectiveness of appropriate dataset selection for low-resource language learning.
On the other hand, the superiority of the PLM is significantly evident, when our methodology incorporates it. As it enhances phoneme sequence decoding by extracting phoneme information more effectively, its performance is amplified from phonologically similar languages.

\textbf{\textit{Comparison with SSL.}} \quad
Recent studies\,\cite{zhao2022improving, ogunremi2023multilingual, chen2024meta} highlight the strong performance of SSL models in low-resource speech recognition.
We fine-tune XLSR-53\,\cite{conneau2020unsupervised}, a Wav2Vec 2.0-based model \cite{baevski2020wav2vec} pre-trained on 56k-hour data across 53 languages.
The fine-tuned SSL model generally shows good performance. However, it struggles in scenarios with very limited training data, such as for Tigrinya.
In contrast, the Conformer-based model trained with our source selection strategy shows superior performance consistently across all languages in terms of average PER. Leveraging languages with high phonological similarity, regardless of their language family, leads to a 55.6\% relative performance gain over the monolingual baseline. This result even dominates the performance of the large-scale SSL model by 21.4\% relatively.
Given that XLSR-53 requires massive amounts of pre-training data and parameters (300M vs. 43M) than the Conformer-based model, this result demonstrates the advantages of our source selection method.

\section{Conclusion}

This study explores cross-lingual phoneme recognition, emphasizing the effect of language similarity on performance in low-resource settings. We compare two primary approaches: the Conformer-based model and the SSL-finetuned model. Specifically, we examine the performance of the Conformer-based model using language families and contrast it with its performance utilizing phonological similarity across the entire language set. Family-based results show that multilingual training is generally more effective with similar languages within the same family, such as Turkic languages. However, linguistic dissimilarity can diminish these benefits. In less similar families, like Indo-Iranian and Afro-Asiatic, same-family multilingual training can reduce performance, but incorporating languages from other families may enhance results. On the other hand, training on phonologically similar languages, regardless of their language family, consistently yields high performance. While SSL typically outperforms monolingual training, multilingual training with closely related languages remains a strong alternative. This comprehensive approach enhances understanding of language selection, while offering valuable insights into the implications of phonetic similarity in cross-lingual contexts.

\vfill
\clearpage
\newpage
\pagebreak

\bibliographystyle{IEEEtran}
\bibliography{IEEEabrv, refs}

\end{document}